\documentclass[english,conference]{IEEEtran}

\usepackage[style=ieee, doi=false, isbn=false]{biblatex}
\usepackage{hyperref}
\usepackage{bm,bbm}
\usepackage{amsthm,amssymb,amsmath,mathtools}
\usepackage{stmaryrd}
\usepackage{microtype}
\usepackage{algorithmic}
\usepackage{xcolor,graphicx,tabularx,float}
\usepackage{babel,csquotes}
\usepackage{ifluatex}
\usepackage{comment}

\makeatletter

\theoremstyle{plain}
\newtheorem{theorem}{Theorem}
\newtheorem{definition}[theorem]{Definition}
\newtheorem{lemma}[theorem]{Lemma}
\newtheorem{proposition}[theorem]{Proposition}

\bmdefine{\bA}{A}
\bmdefine{\ba}{a}
\bmdefine{\bB}{B}
\bmdefine{\bb}{b}
\bmdefine{\bC}{C}
\bmdefine{\bc}{c}
\bmdefine{\bD}{D}
\bmdefine{\bd}{d}
\bmdefine{\bE}{E}
\bmdefine{\be}{e}
\bmdefine{\bF}{F}
\bmdefine{\bf}{f}
\bmdefine{\bG}{G}
\bmdefine{\bg}{g}
\bmdefine{\bH}{H}
\bmdefine{\bh}{h}
\bmdefine{\bI}{I}
\bmdefine{\bi}{i}
\bmdefine{\bJ}{J}
\bmdefine{\bj}{j}
\bmdefine{\bK}{K}
\bmdefine{\bk}{k}
\bmdefine{\bL}{L}
\bmdefine{\bl}{l}
\bmdefine{\bM}{M}
\bmdefine{\bmm}{m}
\bmdefine{\bN}{N}
\bmdefine{\bn}{n}
\bmdefine{\bO}{O}
\bmdefine{\bo}{o}
\bmdefine{\bP}{P}
\bmdefine{\bp}{p}
\bmdefine{\bQ}{Q}
\bmdefine{\bq}{q}
\bmdefine{\bR}{R}
\bmdefine{\br}{r}
\bmdefine{\bS}{S}
\bmdefine{\bs}{s}
\bmdefine{\bT}{T}
\bmdefine{\bt}{t}
\bmdefine{\bU}{U}
\bmdefine{\bu}{u}
\bmdefine{\bV}{V}
\bmdefine{\bv}{v}
\bmdefine{\bW}{W}
\bmdefine{\bw}{w}
\bmdefine{\bX}{X}
\bmdefine{\bx}{x}
\bmdefine{\bY}{Y}
\bmdefine{\by}{y}
\bmdefine{\bZ}{Z}
\bmdefine{\bz}{z}

\bmdefine{\balpha}{\alpha}
\bmdefine{\bbeta}{\beta}
\bmdefine{\bgamma}{\gamma}
\bmdefine{\bGamma}{\Gamma}
\bmdefine{\bdelta}{\delta}
\bmdefine{\bDelta}{\Delta}
\bmdefine{\bepsilon}{\epsilon}
\bmdefine{\bvarepsilon}{\varepsilon}
\bmdefine{\bzeta}{\zeta}
\bmdefine{\bmeta}{\eta}
\bmdefine{\btheta}{\theta}
\bmdefine{\bTheta}{\Theta}
\bmdefine{\biota}{\iota}
\bmdefine{\bkappa}{\kappa}
\bmdefine{\blambda}{\lambda}
\bmdefine{\bLambda}{\Lambda}
\bmdefine{\bmu}{\mu}
\bmdefine{\bnu}{\nu}
\bmdefine{\bpi}{\pi}
\bmdefine{\bPi}{\Pi}
\bmdefine{\brho}{\rho}
\bmdefine{\bsigma}{\sigma}
\bmdefine{\bSigma}{\Sigma}
\bmdefine{\btau}{\tau}
\bmdefine{\bupsilon}{\upsilon}
\bmdefine{\bUpsilon}{\Upsilon}
\bmdefine{\bphi}{\phi}
\bmdefine{\bPhi}{\Phi}
\bmdefine{\bchi}{\chi}
\bmdefine{\bpsi}{\psi}
\bmdefine{\bPsi}{\Psi}
\bmdefine{\bomega}{\omega}
\bmdefine{\bOmage}{\Omega}

\newcommand{\cE}{{\mathcal{E}}}

\newcommand{\cN}{\mathcal{N}}
\newcommand{\cP}{{\mathcal{P}}}

\newcommand{\cX}{\mathcal{X}}

\newcommand{\bbT}{\mathbb{T}}
\newcommand{\bbR}{\mathbb{R}}
\newcommand{\bbC}{\mathbb{C}}
\newcommand{\bbZ}{\mathbb{Z}}

\newcommand{\sT}{{\mathsf{T}}}

\newcommand{\sH}{{\mathsf{H}}}

\DeclareMathOperator{\diag}{diag}

\DeclareMathOperator{\CRLB}{CRLB}

\let\hat\widehat
\let\tilde\widetilde

\newcommand{\pprime}{{\prime\prime}}

\newcommand{\ellp}{{\ell^\prime}}

\makeatother

\IEEEoverridecommandlockouts

\allowdisplaybreaks

\hypersetup{unicode = true, colorlinks,linkcolor=red,anchorcolor=blue,citecolor=blue,
pdftitle = {On the Stability of Super-Resolution and a Beurling--Selberg Type Extremal Problem},
pdfauthor = {Maxime Ferreira Da Costa and Urbashi Mitra}}

\title{On the Stability of Super-Resolution and \\a Beurling--Selberg Type Extremal Problem}

\author{Maxime Ferreira Da Costa and Urbashi Mitra
	\thanks{This work has been funded in part by one or more of the following: Cisco Foundation 1980393, ONR N00014--15--1--2550, ONR 503400--78050, NSF CCF--1410009, NSF CCF--1817200, NSF CCF--2008927, Swedish Research Council 2018--04359, ARO W911NF1910269, and DOE DE--SC0021417.}
	\thanks{Authors' emails: \texttt{\{{mferreira, ubli}\}@usc.edu}}\\
	Department of Electrical and Computer Engineering, University of Southern California, USA
}

\date{\today}

\begin{document}

\maketitle

\begin{abstract}
Super-resolution estimation is the problem of recovering a stream of spikes (point sources) from the noisy observation of a few numbers of its first trigonometric moments.
The performance of super-resolution is recognized to be intimately related to the separation between the spikes to recover. A novel notion of stability of the Fisher information matrix (FIM) of the super-resolution problem is introduced when the minimal eigenvalue of the FIM is \emph{not} asymptotically vanishing. The regime where the minimal separation is inversely proportional to the number of acquired moments is considered. It is shown that there is a separation threshold above which the eigenvalues of the FIM can be bounded by a quantity that does not depend on the number of moments. The proof relies on characterizing the connection between the stability of the FIM and a generalization of the Beurling--Selberg box approximation problem.
\end{abstract}

\section{Introduction}

The classical formulation of super-resolution consists of recovering a stream of point sources (or spikes), characterized by their amplitudes and positions from noisy and distorted measurements. This popular setup serves as a parametric model for many inverse problems in applied
and experimental sciences, such as spectrum analysis, system identification, radar, sonar, and optical imaging~\cite{lindberg_mathematical_2012,lee2012counting}, as well as wireless communications and sensing systems. When the distortion is modeled by a low-pass, bandlimited, shift-invariant point spread function (PSF), the super-resolution problem is also referred to as \emph{line spectral estimation}, considered herein.

While super-resolution has been persistently studied from both theoretical and algorithmic perspectives for several decades, many open related statistical challenges remain. In particular, a complete stability analysis in the presence of noise has not occurred.  Herein, the study of the stability of the Fisher information matrix (FIM) of the super-resolution problem, which is defined by the non-vanishing asymptotic of its smallest eigenvalue when the problem dimension becomes large, is undertaken. The existence of a minimal separation between the sources above which the FIM is stable is established. Through the Cramér--Rao lower bound (CRLB), a new algorithm-free statistical resolution limit to stably recover the point sources is provided. The proof relies on relating the extremal singular values of some generalized Vandermonde matrices, with a novel extension of the Beurling--Selberg box approximation problem.

\subsection{Contributions and Organization of the Paper}

Section~\ref{sec:stabilitySR} describes the signal model and defines the super-resolution model that would be considered in this work. After a review of the existing statistical and algorithmic resolution bounds, the notion of stability of the FIM is presented in Definition~\ref{def:stability} and connected to the CRLB.
The main result of this work is that the stability of the FIM for super-resolution is guaranteed whenever the separation parameter of the problem is greater than $3.54$. This result is presented in  Proposition~\ref{prop:stableSR}.  The key technical result relies on the bounding of the extremal singular values of the sensitivity matrix of the problem.  Further, relevant bounds are established in Lemma~\ref{lem:boundSingularValues} as a function of the separation between the point sources.

The remainder of the paper provides the proof of Lemma~\ref{lem:boundSingularValues}. In Section~\ref{sec:bandlimitedApproximation}, a connection between the Beurling--Selberg box approximation problem and the condition number of Vandermonde matrices with nodes on the unit circle is recalled from the literature. Based on this observation, a novel box approximation problem by bandlimited functions is defined, and a relationship between the solution of this problem and the conditioning of the sensitivity matrix is given in Proposition~\ref{prop:boundsViaExtremalFunctions}. A pair of sub-optimal solutions to the box approximation problem is constructed whenever the separation parameter is large enough, yielding the numerical bounds of Proposition~\ref{prop:stableSR}. A detailed proof of Proposition~\ref{prop:boundsViaExtremalFunctions} is given in Section~\ref{sec:proofOfBoundsViaExtremalFunctions}. Conclusions and future works are drawn in Section~\ref{sec:conclusion}.

\subsection{Notation and Definitions}
Vectors of ${\bbC}^N$ and matrices of $\bbC^{N \times r}$ are denoted by boldface letters $\ba$ and capital boldface letters $\bA$, respectively. The minimal (resp. maximal) eigenvalues and singular values of a matrix $\bA$ are denoted $\lambda_{\min}(\bA)$, $\sigma_{\min}(\bA)$ (resp. $\lambda_{\max}(\bA)$, $\sigma_{\max}(\bA)$). For any function $g\in L_1(\bbR)$, we denote by $\widehat{g}$ its \emph{continuous time Fourier transform}, defined as
\begin{equation}
	\widehat{g}(u) = \int_{\mathbb{R}} g(t) e^{-i 2 \pi u t} dt, \quad \forall u \in \bbR.
\end{equation}
A function $g\in L_1(\bbR)$ is said to be $\beta$-\emph{bandlimited} if for all $u$ such that $\left\vert u \right\vert > \beta$, we have $\widehat{g}(u) = 0$.
For any $\alpha > 0 $, $I_{\alpha}$ is the indicator function of the interval $[-\alpha/2, \alpha/2]$, \emph{i.e.},
\begin{equation}
	I_{\alpha}(t) = \begin{cases}
		1& \mbox{if } \left\vert t \right\vert \leq \alpha/2, \\
		0& \mbox{otherwise.}
	\end{cases}
\end{equation}
which we call the \emph{box function}. We let $\bbT = \bbR \slash \bbZ$ to be the \emph{unidimensional torus}. Let $N = 2n+1$ be an odd number, then for any $\tau \in \bbT$, we denote by $\bv_0(\tau) \in \bbC^N$ and $\bv_1(\tau) = \frac{\mathrm{d}\bv_0(\tau)}{\mathrm{d}\tau} /{\left\lVert \frac{\mathrm{d}\bv_0(\tau)}{\mathrm{d}\tau} \right\rVert_2} \in  \bbC^N$ the unit norm vectors
\begin{subequations}
	\begin{align}
		\bv_0(\tau) ={}& \frac{1}{\sqrt{N}}\left[e^{- i 2 \pi (-n) \tau}, \dots, e^{- i 2 \pi n \tau}\right]^\top \\
		\bv_1(\tau) ={}& \frac{C_N}{\sqrt{N}} \left[-i 2 \pi (-n) e^{- i 2 \pi (-n) \tau}, \dots, -i 2 \pi n e^{- i 2 \pi n \tau}\right]^\top,
	\end{align}
\end{subequations}
where $C_N = \sqrt{\frac{3}{\pi^2 (N-1) (N+1)}}$ is a normalization constant.
For any vector $\btau = {[\tau_1, \dots, \tau_r]}^{\sT} \in {\bbT}^r$, we define by $\bV_0(\btau), \bV_1(\btau) \in \bbC^{N \times r}$, the \emph{generalized Vandermonde matrices}
\begin{subequations}
	\begin{align}
		\bV_0(\btau) ={}& \left[ \bv_0(\tau_1), \dots, \bv_0(\tau_r) \right], \\
		\bV_1(\btau) ={}& \left[ \bv_1(\tau_1), \dots, \bv_1(\tau_r) \right].
	\end{align}
\end{subequations}
The concatenation $\bW(\btau) = [\bV_0(\btau), \bV_1(\btau)] \in \bbC^{N \times 2r}$ is refereed to as the \emph{sensitivity matrix}. Finally, the \textit{wrap-around distance} $\Delta(\btau)$ is defined by the minimal distance between two pairs of points in $\btau$ over the torus,  \textit{i.e.}
\(		\Delta(\btau) \triangleq \min_{\ell \neq \ell^\prime} \inf_{j \in \bbZ} \left\vert \tau_\ell - \tau_{\ell^\prime} + j \right\vert
\).

\section{The Stability of Super-Resolution}\label{sec:stabilitySR}

\subsection{Signal Model}
Consider $r$ sources characterized by a vector $\bc = {[c_1, \dots, c_r]}^{\sT} \in \bbC^r$ of $r$ complex amplitudes, and a vector $\btau = {[\tau_1, \dots, \tau_r]}^{\sT} \in \bbT^r$. Our super-resolution problem is to estimate the $2r$ parameters $\{\bc, \btau\}$ of the continuous time domain signal
\( \label{eq:timeDomainSignal}
	x(t) = \sum_{\ell=1}^r c_\ell \delta(t-\tau_\ell)
\)
from the noisy observation $\by =[y_{-n}, \dots, y_0,\dots y_n] \in \mathbb{C}^{N}$ of the form
\begin{align}
	\by ={}& { \left[ \hat{x}(-n), \dots, \hat{x}(n) \right] }^{\sT} + \bw \nonumber  \\
	={}& \sum_{\ell=1}^r c_\ell \bv_0(\tau_\ell) + \bw = \bV_0(\btau) \bc + \bw, \label{eq:observationModel}
\end{align}
where $\bw \sim \cN\left(\bm{0}, \sigma^2 \bI_N \right)$ is assumed to be white Gaussian noise with variance $\sigma^2$, and $r \leq n$ to ensure the uniqueness of the solution with $r$ components in absence of noise~\cite{fuchs_sparsity_2005}. Up to a scaling, it is assumed that $1 \leq \left\vert c_\ell \right\vert \leq \kappa$ of any $\ell=1,\dots,r$, where $\kappa$ is the \emph{dynamic range} of the problem. As the vectors $\bc$ and $\btau$ are of different units, and that the statistical error of an estimator of $\tau$ is expected to be inversely proportional to the number of measurement $N$, we denote by $\tilde{\btau} = C_N^{-1} \btau$ the normalized location of the points. We seek to recover, without loss of generality, the set of parameters  ${\btheta} = \{ \bc, \tilde{\btau} \}$.

\subsection{Stability of Super-Resolution and Resolution Limits} \label{subsec:stability}

\begin{figure}[t]
	\centering
	\includegraphics[width=0.95\columnwidth]{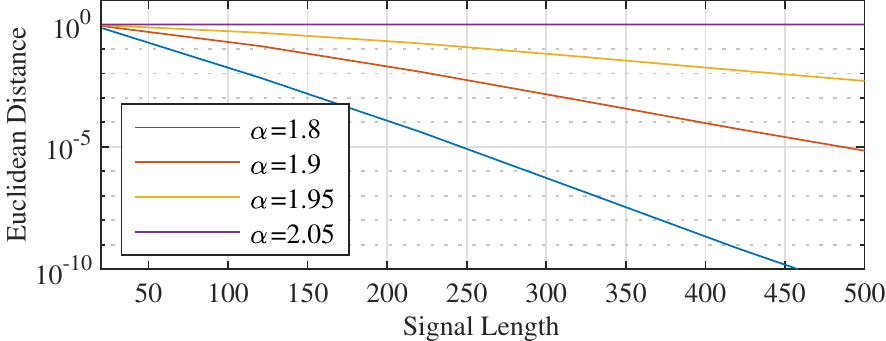}
	\vspace*{-0.05in}
	\caption{Euclidean distance $\left\Vert \bx_{1} - \bx_{2} \right\Vert_2$ between two distinct noiseless signals with separation parameter $\alpha$ and $r = N/6$ sources, for different values of $\alpha$ and of the signal length $N$. Herein, the minimal cross-separation between the sets of locations $\btau_1$ and $\btau_2$ composing $\bx_1$ and $\bx_{2}$  is set to $\alpha/2$.}
	\label{fig:resolutionLimit}
\end{figure}

An important statistical analysis goal for super-resolution is to guarantee that an estimator $\widehat{\btheta}$ remains reasonably close to the ground truth $\btheta$. A variety of definitions exists in the literature to characterize the stability of super-resolution. All have been shown to be related to the separation parameter $\alpha \triangleq N\Delta(\btau)$ between the sources, as empirically established by Rayleigh  (see \emph{e.g.}~\cite{born2013principles}). We give a brief review of the key prior results in this context. When the sources are on a discrete grid, the stability of super-resolution has been shown to be related to the Rayleigh index of the support set~\cite{donoho1992superresolution, demanet2015recoverability}. In the last decade, there has been a resurgence in interest in characterizing the resolution limit for sources lying \emph{off-the-grid}. It is well-established that some instances of super-resolution are statistically unidentifiable under a fixed noise level $\sigma^2$ when $N$ grows large if the separation parameter satisfies $\alpha < 2$. That is, two ill-separated sets of distinct parameters can lead to arbitrarily close signals. This phenomena is highlighted in  Figure~\ref{fig:resolutionLimit} for the construction of two signals with $r=N/6$ regularly spaced sources, and can be explained by a phase transition in the conditioning of the matrix $\bV_0(\btau)$ whenever $\alpha < 1$~\cite{aubel_theory_2018,moitra_super-resolution_2015}. Additional studies of this property are proposed in the context of colliding pairs of sources~\cite{batenkov2021super, kunis2021condition, diederichs_well-posedness_2019}, and the impact of the bandwidth selection was highlighted in \cite{batenkov2019rethinking}.

Other lines of work aim to characterize instead the resolution limits of specific estimators for the super-resolution problem. For instance, stability guarantees for the MUSIC algorithm are provided as a function of the separation parameter in~\cite{liao2016music} in the noisy setting. The atomic norm denoiser~\cite{candes_towards_2014,de_castro_exact_2012,chi_harnessing_2019} (\emph{a.k.a} Beurling-LASSO) can fail to recover $\btheta$ in the absence of noise whenever $\alpha < 2$~\cite{ferreira_da_costa_tight_2018}, and is guaranteed to succeed for $\alpha \geq 2.56$~\cite{candes_towards_2014,fernandez-granda_super-resolution_2016}. Stability and denoising performance under Gaussian white noise are given for a separation $\alpha \geq 5.01$ in~\cite{li_approximate_2018}. Finally, the support stability~\cite{duval_exact_2015}, ensuring that the estimate $\widehat{\btheta}$ as the same number of components as of the ground truth $\btheta$, is  established for $\alpha \geq 2.27$ in~\cite{Ferreira2020Stable}.

\subsection{Stability of the Fisher Information Matrix}

In this paper, we associate the stability of super-resolution with the property of the Fisher information matrix $\bJ(\btheta)$ that its smallest eigenvalue is strictly bounded away from $0$. A formal definition of stability is given in the following.
\begin{definition}[Stability of the Fisher Information Matrix]\label{def:stability}
	The super-resolution problem is to be \emph{stable} for a separation parameter  $\alpha$ and a dynamic range $\kappa$ if and only if there exists a constant $C(\alpha,\kappa)$ independent of $N$ such that for any set of parameters $\btheta$ with $N \Delta(\btau) \geq \alpha$ and $1 \leq \min\{\bc \} \leq \max\{\bc \} \leq \kappa$
	\begin{equation}
		\lambda_{\min} \left( \bJ(\btheta) \right) \geq \sigma^{-2} C(\alpha, \kappa).
	\end{equation}
\end{definition}

When read in the light of the CRLB, which states that the covariance $\bSigma_{\hat{\btheta}}$ of an unbiased estimator $\widehat{\btheta}$ of $\btheta$ given $\by$ is such that $\bSigma_{\widehat{\btheta}} - \bJ(\btheta)^{-1}$ is a positive semi-definite Hermitian matrix (see \emph{e.g.}~\cite{scharf1993geometry,smith2005statistical} ), Definition~\ref{def:stability} can be reinterpreted in terms of the stability of estimates of linear forms as follows.
\begin{proposition}[Bounded CRLB Property]\label{prop:CRLB}
The super-resolution problem is stable in the sense of Definition~\ref{def:stability} if and only if the CRLB of any linear form $z(\btheta)$ on the parameters $\btheta$ of the kind $z(\btheta) = \bq^{\sH} \btheta$ verifies
    \begin{equation}
        \CRLB(z(\btheta);\by) \leq \sigma^2 C(\alpha,\kappa)^{-1} \left\Vert \bq \right\Vert_2^2,
    \end{equation}
    independently of $N$.
\end{proposition}

The contrapositive of Proposition~\ref{prop:CRLB} induces that the super-resolution problem is \emph{not} stable if and only if there exists a linear form on the parameters that cannot be reliably estimated in the limit $N \to \infty$. Hence, Definition~\ref{def:stability} is stronger than requiring the stability of each individual parameter composing $\btheta$, which is commonly adopted in the literature. Yet, the notion of stability considered in this work doesn't guarantee the actual existence of a consistent estimator of a given linear form. The main result of this paper, in the following proposition, establishes stability in the sense of the Fisher information provided that the separation parameter $\alpha$ is sufficiently large.

\begin{proposition}[Sufficient Condition for Stability]\label{prop:stableSR}
	Suppose that $\alpha \geq 3.54$, then under Gaussian white noise $\bw \sim \cN \left(\bm{0}, \sigma^2 \bI \right)$, and for any $\kappa \geq 1$ the super-resolution problem is stable in the sense of Definition~\ref{def:stability}. Moreover, there exist a non-decreasing function $h_{-}(\alpha) > 0$ and a non-increasing function $h_{+}(\alpha) >0$ such that
	\begin{subequations}
		\begin{align}
			\lambda_{\min} \left( \bJ(\btheta) \right) \geq{}& \sigma^{-2} h_{-}(\alpha) > 0 \\
			\lambda_{\max} \left( \bJ(\btheta) \right) \leq{}& \sigma^{-2} \kappa^2 h_{+}(\alpha).
		\end{align}
	\end{subequations}
\end{proposition}

Figure~\ref{fig:extremalValues} compares the empirical realizations of the extremal eigenvalues of the FIM with the functions $h_-(\cdot)$ and $h_+(\cdot)$ that will be later constructed in Section~\ref{sec:bandlimitedApproximation}. The evaluation $h_-(3.54) > 0$ suggests that $\alpha > 3.54$ is \emph{sufficient} to guarantee the stability  in the sense of Definition~\ref{def:stability}, however there is no reason to believe this bound will be tight. Instead, the numerical experiments  suggest a phase transition on $\lambda_{\min} \left( \bJ(\btheta) \right)$ whenever $\alpha < 2$ as the eigenvalue rapidly vanishes to $0$. This observation also coincides with the limit discussed in Section~\ref{subsec:stability}. We leave a sharpening of the bounds for future work. The sequel is devoted to proving Proposition~\ref{prop:stableSR}.

\subsection{Proof of Proposition~\ref{prop:stableSR}}

We start the analysis by deriving the FIM $\bJ(\btheta)$ (see \emph{e.g.}~\cite{scharf1993geometry,pakrooh2015analysis} of the super-resolution problem~\eqref{eq:observationModel}
\begin{equation}\label{eq:expressionFIM}
	\bJ(\btheta) = \sigma^{-2} \diag(\bm{1},\bc)^{\sH} \bW(\btau)^{\sH} \bW(\btau)  \diag(\bm{1},\bc),
\end{equation}
where $\diag(\bm{1},\bc)$ denotes the $2r \times 2r$ diagonal matrix whose first $r$ diagonal entries are equal to $1$ and the following $r$ are equal to $\bc$.  Some properties of this matrix where studied in \cite{li2000parametric,blu2008sparse}. Equation~\eqref{eq:expressionFIM} immediately yields that $\bJ(\btheta)$ is a positive semi-definite matrix with the following bounds on its extremal eigenvalues
\begin{subequations}\label{eq:boundEigenvalues}
	\begin{align}
		\lambda_{\min} \left( \bJ(\btheta) \right) \geq{}& \sigma^{-2} \min(1,\left\vert c_1 \right\vert,\dots, \left\vert c_r \right\vert)^2 \sigma_{\min} \left(\bW(\btau) \right)^2 \nonumber\\
		\geq{}& \sigma^{-2} \sigma_{\min} \left(\bW(\btau) \right)^2\\
		\lambda_{\max} \left( \bJ(\btheta) \right) \leq{}& \sigma^{-2} \max(1,\left\vert c_1 \right\vert,\dots, \left\vert c_r \right\vert)^2 \sigma_{\max} \left(\bW(\btau) \right)^2 \nonumber\\
		\leq{}& \sigma^{-2}\kappa^2  \sigma_{\max} \left(\bW(\btau) \right)^2.
	\end{align}
\end{subequations}
Therefore, the crux of the problem consists of ensuring the existence of the two functions $h_-(\cdot)$ and $h_+(\cdot)$, satifying the conditions stated in Proposition~\ref{prop:stableSR} and that minorize (resp. majorize) the quantities $\sigma_{\min} \left(\bW(\btau) \right)^2$ (resp. $\sigma_{\max} \left(\bW(\btau) \right)^2$).  This goal is accomplished in the following lemma and proved in Section~\ref{sec:bandlimitedApproximation} after a detour on the Beurling--Selberg extremal approximation problem.

\begin{lemma}[Conditioning of $\bW(\btau)$] \label{lem:boundSingularValues}
	 There exist a non-decreasing function $h_{-}(\cdot)$ and a non-increasing function $h_{+}(\cdot)$ such that for any $N$ and any $\btau \in {\bbT}^r$ verifying $3.54 \leq \alpha \leq N \Delta(\btau)$, we have the following inequalities
	\begin{subequations}
		\begin{align}
			\sigma_{\min} \left(\bW(\btau) \right)^2 \geq{}& h_-(\alpha) > 0 \\
			\sigma_{\max} \left(\bW(\btau) \right)^2 \leq{}& h_+(\alpha).
		\end{align}
	\end{subequations}
\end{lemma}
The proof of Proposition~\ref{prop:stableSR} immediately follows from~\eqref{eq:boundEigenvalues} and Lemma~\ref{lem:boundSingularValues}. \qed

\section{Relationship with Bandlimited Approximation}\label{sec:bandlimitedApproximation}

\subsection{The Beurling--Selberg Box Approximation Problem} \label{subsec:Beurling--Selberg_approx}

\begin{figure}[t]
	\centering
	\includegraphics[width=0.95\columnwidth]{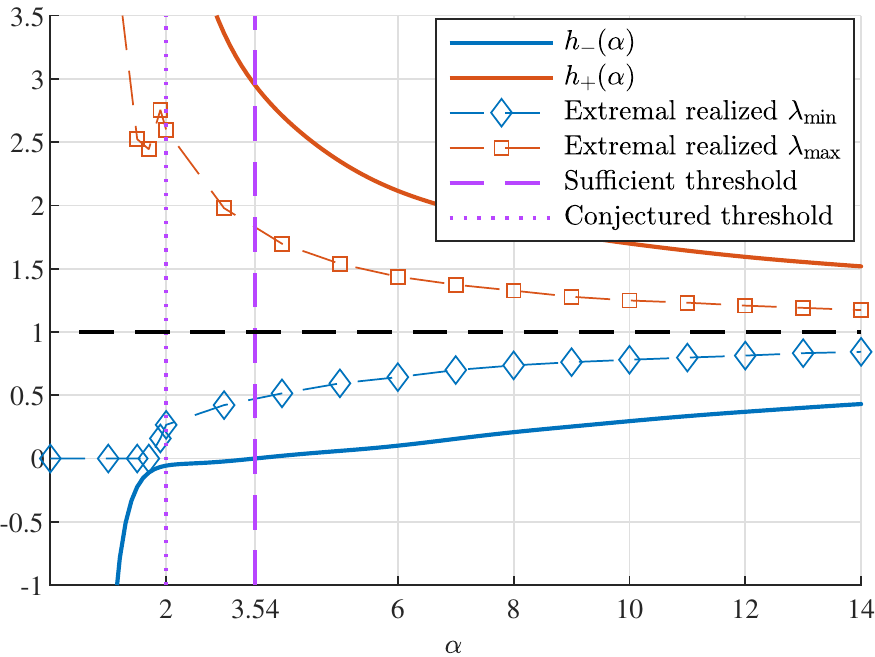}
	\caption{The theoretical bounds $h_-(\alpha)$ and $h_+(\alpha)$ versus the experimental extremal realizations of $	\lambda_{\min} \left( \bJ(\btheta) \right)$ and $	\lambda_{\max} \left( \bJ(\btheta) \right)$ when $N\Delta(\btau) = \alpha$. Parameter values are $N=1001$, $\kappa = 1$ and $\sigma^2 = 1$. Experimental results are computed over 200 randomized trials.}
	\label{fig:extremalValues}
\end{figure}

The Beurling--Selberg extremal box approximation problem consists of finding a minorant and a majorant to the box function $I_\alpha$ that are 1-bandlimited while minimizing the $L_1(\bbR)$ distance to $I_\alpha$~\cite{selberg2014collected,akhilesh2013remark}. Selberg proposed a construction of two approximation functions $B_-^{\alpha}(\cdot)$ and $B_+^{\alpha}(\cdot)$ by leveraging the properties of Beurling's extremal approximant of the signum function~\cite{vaaler1985some}. In particular, those two functions are said to be \emph{extremal} in the sense that they achieve \(	\widehat{B}_-^{\alpha}(0) = 1 - \alpha^{-1}\)  and \(	\widehat{B}_+^{\alpha}(0) = 1 + \alpha^{-1} \)
and that any 1-bandlimited minorant $F_-^\alpha$ (resp. majorant $F_+^\alpha$) of $I_\alpha$ satisfies $\widehat{F}_-(0) \leq 1 - \alpha^{-1} $ (resp. $\widehat{F}_+(0) \geq 1 + \alpha^{-1}$). Those properties are particularly interesting because they can be used to bound, in a fairly elegant manner, the singular values of Vandermonde matrices $\bV_0(\btau)$ for any $\btau$ with separation parameter $\alpha = N \Delta(\btau)$ through (see \emph{e.g.}~\cite{aubel2019vandermonde})
\begin{subequations} \label{eq:bound_V0}
	\begin{align}
		\sigma_{\min}(\bV_0(\btau))^2 \geq{}& 1 - \alpha^{-1} \label{eq:bound_min_V0}\\
		\sigma_{\max}(\bV_0(\btau))^2 \leq{}& 1 + \alpha^{-1}.
	\end{align}
\end{subequations}
Henceforth, given a fixed $\alpha>1$, it is immediately inferred from~\eqref{eq:bound_V0} that $\sigma_{\min}(\bV_0(\btau))^2$  is non-vanishing, and that $\sigma_{\max}(\bV_0(\btau))^2$ is upper bounded by a quantity independent of $N$ for any values of $\alpha > 0$.

\subsection{Higher Order Box Approximations} \label{subsec:existence}

With the purpose of demonstrating Lemma~\ref{lem:boundSingularValues}, we introduce a novel \emph{higher order} box approximation problem, which imposes an additional assumption on the decay rate of the minorizing and majorizing functions of $I_\alpha$. The minorant set $\cE_\alpha^-$ and minorant set $\cE_\alpha^+$ are defined as follows:
\begin{definition}[2nd order minorant and majorant set]\label{def:approximationSets}~\linebreak
\underline{Minorant set:} Given $\alpha > 0$, we let $\cE_\alpha^- \subset L_1(\bbR)$ the sets of functions $\mathcal{X}_- \in L_1(\mathbb{R})$ satisfying the following properties:
\begin{enumerate}
	\item $\mathcal{X}_-$ is $1$-bandlimited;
	\item $\mathcal{X}_-$ minorizes $I_\alpha$, i.e. $\mathcal{X}_-(t) \leq{} I_{\alpha}(t)$ for all $t$;
	\item $\cX_-(t) = \mathcal{O}(t^{-(3+\varepsilon)})$ when $\left\vert t \right\vert \to \infty$ for some $\varepsilon > 0$;
    \item $\widehat{\cX}_-^\prime(0) ={} 0$.
\end{enumerate}
\underline{Majorant set:} Given $\alpha > 0$, we let $\cE_\alpha^+ \subset L_1(\bbR)$ the sets of functions $\mathcal{X}_+ \in L_1(\mathbb{R})$ satisfying the following properties:
\begin{enumerate}
	\item $\mathcal{X}_+$ is $1$-bandlimited;
	\item $\mathcal{X}_+$ majorizes $I_\alpha$, i.e. $\mathcal{X}_+(t) \geq{} I_{\alpha}(t)$ for all $t$;
	\item $\cX_+(t) = \mathcal{O}(t^{-(3+\varepsilon)})$ when $\left\vert t \right\vert \to \infty$ for some $\varepsilon > 0$;
	\item $\widehat{\cX}_+^\prime(0) ={} 0$.
\end{enumerate}
\end{definition}

We note that, by the Riemann--Lebesgue lemma, the third condition in both cases guarantees the Fourier transform of the functions in $\cE_\alpha^-$ and $\cE_\alpha^+$ to be at least twice differentiable, which is needed in the fourth condition. The sets proposed in Definition~\ref{def:approximationSets}  only differ from the original Beurling--Selberg approximation problem in the third and fourth assumptions. However, it can be easily checked from their definitions~\cite{vaaler1985some} that $B_\alpha^-(\cdot)$ and $B_\alpha^+(\cdot)$ do not belong to $\cE_\alpha^-$ and $\cE_\alpha^+$, respectively, as they decay at an asymptotic rate $B_\alpha^-(t) \sim B_\alpha^+(t) \sim C t^{-2}$ when $\left\vert t \right\vert \to \infty$, thus fail to meet the third assumption.

The following proposition is at the core of the proof of Proposition~\ref{prop:stableSR} and establishes a relationship between the functions belonging to the set $\cE_\alpha^-$ and $\cE_\alpha^+$ and the extremal singular values of the matrix $\bW(\btau)$.

\begin{proposition}[Conditioning of $\bW(\btau)$ via Bandlimited Functions]\label{prop:boundsViaExtremalFunctions}
	Let $N = 2n+1$ be an odd integer. For any fixed $\alpha > 0$, let $\btau \in {\bbT}^r$, be such that $N \Delta(\btau) \geq \alpha$. Then, for any function $ \cX_- \in \cE_\alpha^-$ and $ \cX_+ \in \cE_\alpha^+$,  we have
	\begin{subequations}\label{eq:sigmaBounds}
		\begin{align}
			{\sigma_{\min} (\bW(\btau))}^2 \geq{}&
			\min \left\{ \alpha^{-1} \widehat{\cX}_-(0), - \frac{3}{\pi^2} \alpha^{-3} \widehat{\cX}_-^{\pprime}(0) \right\} \label{eq:lowerBound} \\
			{\sigma_{\max}(\bW(\btau))}^2 \leq{} &
			\max \left\{  \alpha^{-1} \widehat{\cX}_+(0), - \frac{3}{\pi^2} \alpha^{-3} \widehat{\cX}_+^{\pprime}(0)  \right\}.\label{eq:upperBound}
		\end{align}
	\end{subequations}
\end{proposition}
A few remarks are of order regarding the statement of Proposition~\ref{prop:boundsViaExtremalFunctions}. The crux of the problem is two fold. First, the extremal singular values can be controlled given the existence of functions within the the sets $\cE_\alpha^-$ and $\cE_\alpha^+$, hence a strategy to prove Proposition~\ref{prop:stableSR} is to show that those two sets are not empty for a large enough value of $\alpha$. Second, noticing that $\widehat{I}_\alpha(0) = \int_\mathbb{R} I_\alpha(t) dt = \alpha$ and ${\widehat{I}_\alpha}^{\pprime}(0) = \int_\mathbb{R} -4 \pi^2 t^2 I_\alpha(t) dt = -\frac{\pi^2}{3} \alpha^3$, the scaling of the bounds~\eqref{eq:sigmaBounds} can be considered to be tight in the sense that the right sides of both equations with the functions $\cX_- = \cX_+ = I_\alpha$ equal $1$. Hence, better bounds are achieved for functions in $\cE_\alpha^-$ and $\cE_\alpha^+$ that are close to the box function $I_\alpha$. The best bounds on the singular values of the sensitivity matrix $\bW(\btau)$ in Proposition~\ref{prop:boundsViaExtremalFunctions} are provided by the \emph{extremal functions} that achieve the maximal and minimal values on the right-hand sides of equations~\eqref{eq:lowerBound} and~\eqref{eq:upperBound} respectively. However, to our knowledge, there are no known derivations to date for those extremal functions.  We seek to ameliorate that lack in the sequel.

\subsection{Approximation via Polynomial Transforms}

 In this section, we construct a pair of suboptimal functions $G_\alpha^-(\cdot)$ and $G_\alpha^+ (\cdot)$ belonging to the sets $\cE_\alpha^-$ and $\cE_\alpha^+$, respectively and that achieve a reasonable approximation of the box function $I_\alpha$. We proceed by applying a transform $\cP$ to the Beurling--Selberg extremal functions discussed in~\ref{subsec:Beurling--Selberg_approx} which preserves the first and second properties of $\cE_\alpha^-$ and $\cE_\alpha^+$ in Definition~\ref{def:approximationSets} respectively. Mathematically, it is sufficient to ensure that $\cP$ satisfies the following properties:
\begin{enumerate}
	\item For any 1-bandlimited function $F$, then $ \cP\{F\}$ is also 1-bandlimited;
	\item If $F$ minorizes (resp. majorizes) the box function $I_\alpha$ then $ \cP\{F\}$ also minorizes (resp. majorizes) $I_\alpha$;
	\item If $F(t) \to 0$ at $\left\vert t \right\vert \to \infty$ then $\cP \{ F \}(t) = \mathcal{O}(F(t)^2)$ in the limit $\left\vert t \right\vert \to \infty$.
\end{enumerate}
Polynomial transforms are a good fit to achieve those three conditions, since any polynomial $P$ of degree $d$ preserves the spectral support of $F$ through the map $\cP\{F\}(t) = P(F(\frac{t}{d}))$. It can be checked that, while not unique, the polynomial $P(t)=t^3$ is the minimal degree polynomial for which $\cP$ matches those three properties. This suggests a construction of $G_\alpha^-(\cdot)$ and $G_\alpha^+(\cdot)$ as in Lemma~\ref{lem:polynomialTransform}, whose proof is omitted for brevity.

\begin{lemma}\label{lem:polynomialTransform}
	Let $\alpha > 0$, and denote by $G_\alpha^-$ and $G_\alpha^+$ the functions given by
	\begin{subequations}\label{eq:DefinitionOfG}
		\begin{align}
			G_\alpha^-(t) &{}\triangleq \cP \{B_{\alpha/3}^{-} \}(t) = {B_{\alpha/3}^{-}(t/3)}^3 \label{eq:DefinitionOfG_minorant}\\
			G_\alpha^+(t) &{}\triangleq \cP \{B_{\alpha/3}^{+} \}(t) = {B_{\alpha/3}^{+}(t/3)}^3,
		\end{align}
	\end{subequations}
	then we have that $G_\alpha^- \in \cE_-^\alpha$ and $G_\alpha^+ \in \cE_+^\alpha$.
\end{lemma}

\begin{figure}[t]
	\centering
	\includegraphics[trim={0.52in 0in 0.52in 0in},clip,width=\columnwidth]{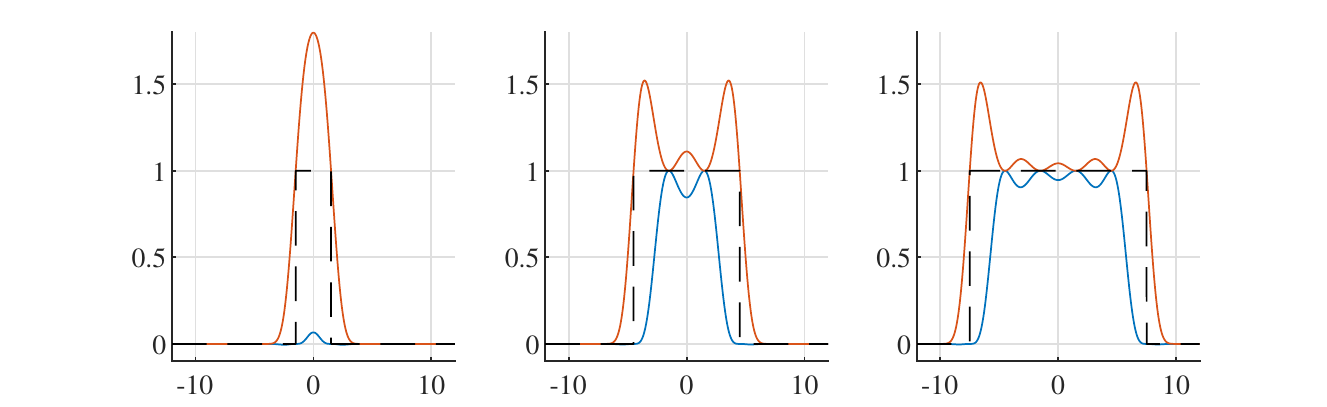}
	\caption{The graph of the functions $G_\alpha^-(t)$ (in blue), $G_\alpha^+(t)$ (in red) and $I_\alpha(t)$ (in black dashed lines). Left: $\alpha = 3$; Middle: $\alpha = 9$; Right: $\alpha=15$}
	\label{fig:polynomialTransform}
\end{figure}

Figure~\ref{fig:polynomialTransform} shows the graph of these functions and their Fourier transforms for three different values of the separation parameter~$\alpha$.

We now have all the elements to prove Lemma~\ref{lem:boundSingularValues} which is integral to the proof of Proposition~\ref{prop:stableSR}. We let $h_-(\alpha)$ and $h_+(\alpha)$ be the value of the right-hand side of the bounds~\eqref{eq:sigmaBounds} applied to the functions $G_\alpha^-(\cdot) \in \cE_\alpha^-$ and $G_\alpha^+(\cdot) \in \cE_\alpha^+$ defined in~\eqref{eq:DefinitionOfG}, respectively.
The numerical evaluation of those quantity shown in Figure~\ref{fig:extremalValues} ensures that $h_-(\alpha) > 0$ whenever $\alpha \geq 3.54$, which concludes on the desired statement. \qed

\section{Proof of Proposition~\ref{prop:boundsViaExtremalFunctions}} \label{sec:proofOfBoundsViaExtremalFunctions}

We focus on proving that~\eqref{eq:lowerBound} holds, as~\eqref{eq:upperBound} can be derived in an analogous manner.
Consider a vector $\btau \in \mathbb{T}^r$ of an arbitrary number $r$ of elements, and let $\alpha = N \Delta(\btau)$. We fix $\bu_0,\bu_1 \in \bbC^r$ and we let by $\bu = [\bu_0^\top, \bu_1^\top]^\top \in \bbC^{2r}$, the concatenation of the two vectors. Assume that the function $\cX_-(\cdot)$ is in the set $\cE_\alpha^-$. We start by introducing the auxiliary functions $\varphi_0(\cdot)$ and $\varphi_1(\cdot)$ defined for all $y \in \mathbb{R}$ by
\begin{subequations}
	\begin{align}
		\varphi_0 (y) ={}& \frac{1}{\sqrt{N}} \sum_{\ell = 1}^r u_{0,\ell} e^{- i 2 \pi \tau_\ell y}\\
		\varphi_1 (y) ={}& -\frac{C_N}{\sqrt{N}} \sum_{\ell = 1}^r u_{1,\ell} i 2 \pi y e^{- i 2 \pi \tau_\ell y}.
	\end{align}
\end{subequations}
and let by  $S_p(\ell,\ellp)$ for $p = 0,1,2$ the series
\begin{equation}
	S_p(\ell,\ellp) ={} \sum_{k= -\infty}^{\infty} \cX_{-} \left( k \Delta( \btau ) \right) \left(i 2\pi k \right)^{p} e^{-i 2\pi k (\tau_\ell - \tau_\ell^\prime)}.
\end{equation}
Given these definitions, we can introduce $\cX_-$ to minorize the quantity $\left\Vert \bW(\btau) \bu \right\Vert_2^2$ as follows
\begin{align} \label{eq:expandSquare}
	\left\Vert \bW(\btau) \bu \right\Vert_2^2 ={}& \sum_{k=-n}^n \left\vert \varphi_0(k) + \varphi_1(k) \right\vert^2 \nonumber\\
	={}& \sum_{k=-\infty}^\infty I_{\alpha}(k \Delta(\btau)) \left\vert \varphi_0(k) + \varphi_1(k) \right\vert^2 \nonumber\\
	\geq{}& \sum_{k=-\infty}^\infty \cX_- \left(k \Delta(\btau) \right) \left\vert \varphi_0(k) + \varphi_1(k) \right\vert^2 \nonumber\\
	={}& \frac{1}{N} \sum_{\ell = 1}^r \sum_{\ell^\prime = 1}^r u_{0,\ell} u_{0,\ell^\prime} S_0(\ell,\ellp) \nonumber\\
	{}& \quad -  \frac{2 C_N}{N} \Re \left( \sum_{\ell = 1}^r \sum_{\ell^\prime = 1}^r u_{0,\ell} u_{1,\ell^\prime} S_1(\ell,\ellp) \right) \nonumber\\
	{}& \quad + \frac{{C_N}^2}{N}  \sum_{\ell = 1}^r \sum_{\ell^\prime = 1}^r u_{1,\ell} u_{1,\ell^\prime} S_2(\ell,\ellp),
\end{align}
where the first inequality stems from the fact that $k \Delta(\btau) \in [-\alpha/2, \alpha/2]$ if and only if $\left\vert k \right\vert \leq n$, and the second because $\cX_-(t) \leq I_\alpha(t)$ for all $t\in \mathbb{R}$ by assumption. Moreover, the terms of the series $S_p(\ell, \ellp)$ are equivalent to the function $h_p: x \mapsto \cX_-(x \Delta(\btau)){(i 2\pi x)}^p e^{-i 2\pi x(\tau_\ell - \tau_\ellp)}$ when the argument is an integer. By the third assumption in Definition~\ref{def:approximationSets}, the functions $h_p$ are absolutely summable for $p=0,1,2$ with continuous time Fourier transform $\widehat{h}_p(u) = \Delta(\btau)^{-(1+p)} \widehat{\cX}_-^{(p)}(\Delta(\btau)^{-1}(u - \tau_\ell + \tau_\ellp))$, which is of finite support by the first assumption and bounded by the third one, therefore also absolutely summable. Hence, the Poisson summation formulae can be applied on the series $S_p(\ell, \ellp)$. They yield for all pairs $(\ell, \ellp) \in \left\{1,\dots,r\right\}^2$ and all $p=0,1,2$
\begin{align}\label{eq:PoissonSummation}
	S_p(\ell, \ellp) = \Delta(\btau)^{-(1+p)} \sum_{j=-\infty}^{\infty} \widehat{\cX}_- ^{(p)} \left( \Delta(\btau)^{-1} (j- \tau_\ell + \tau_{\ell^\prime}) \right).
\end{align}
By our assumption on the separation, we have that $\Delta(\btau)^{-1} (j- \tau_\ell + \tau_{\ell^\prime}) > 1$ for any $j \in \bbZ$ and any pair $(\ell, \ellp)$ unless $\ell = \ellp$ and $j=0$. As $\cX_-(\cdot)$ is $1$-bandlimited, we conclude that the terms on the right hand side of~\eqref{eq:PoissonSummation} are all $0$ unless $\ell = \ellp$ and $j=0$. Thus,~\eqref{eq:PoissonSummation} reduces for $p=0,1,2$ to
\begin{equation}\label{eq:useSeparation}
		S_p(\ell, \ellp) = \begin{cases} \Delta(\btau)^{-(p+1)} \widehat{\cX}_-^{(p)} \left( 0 \right) & \text{if } \ell = \ellp,\\
		0& \text{otherwise}.
		\end{cases}
\end{equation}
Substituting~\eqref{eq:useSeparation} into~\eqref{eq:expandSquare} with $\widehat{\cX}_-^\prime(0) = 0$, $\frac{C_N^2}{N} \geq \frac{3}{\pi N^3}$ for all $N > 1$, and $\alpha = N \Delta(\btau)$ yields
\begin{align}\label{eq:bound1}
	\MoveEqLeft \left\Vert \bW(\btau) \bu \right\Vert_2^2 & \nonumber \\
	\geq{}& \left(N \Delta(\btau)\right)^{-1} \widehat{\cX}_-(0) \left\Vert \bu_0\right\Vert_2^2
	- \frac{{C_N}^2}{N} \Delta(\btau)^{-3} \widehat{\cX}_-^{\pprime}(0) \left\Vert \bu_1\right\Vert_2^2  \nonumber \\
	={}& \alpha^{-1} \widehat{\cX}_-(0) \left\Vert \bu_0\right\Vert_2^2 - \frac{3}{\pi^2} \alpha^{-3} \widehat{\cX}_-^{\pprime}(0) \left\Vert \bu_1\right\Vert_2^2.
\end{align}
Finally, the definition of the singular value and~\eqref{eq:bound1} imply
\begin{align}\label{eq:bound2}
	{\sigma_{\min}(\bW(\btau))}^2 ={}& \min_{\left\Vert \bu\right\Vert_2^2 =1 } \left\Vert \bW(\btau) \bu \right\Vert_2^2 \nonumber \\
	\geq{}& 	\min \left\{  \alpha^{-1} \widehat{\cX}_-(0), - \frac{3}{\pi^2} \alpha^{-3} \widehat{\cX}_-^{\pprime}(0) \right\},
\end{align}
which yields our desired statement. \qed

\section{Conclusion and Future Work} \label{sec:conclusion}

In the present work, we considered the stability of the Fisher information matrix of the super-resolution problem in the regime where the separation between the sources is inversely proportional to the number of measurements. Proposition~\ref{prop:stableSR} establishes the stability of the FIM whenever the separation parameter verifies $\alpha \geq 3.54$. This result was demonstrated by unveiling in Proposition~\ref{prop:boundsViaExtremalFunctions} a connection between the singular values of the sensitivity matrix and the solutions of a novel Beurling--Selberg type approximation problem, for which we derived a  pair of sub-optimal solutions. We leave for a sequel the study of the extremal solutions to this approximation problem in the hope of guaranteeing the stability of the FIM up to the conjectured threshold $\alpha > 2$.

\printbibliography{}

\end{document}